# A Weakly Coordinating Anion Substantially Enhances Carbon Dioxide Fixation by Calcium and Barium Salts

Vitaly V. Chaban,[1,A] Nadezhda A. Andreeva,[2] and Pavel N. Vorontsov-Velyaminov[3]

1) Federal University of São Paulo, São Paulo, Brazil.

2) PRAMO, St Petersburg, Leningrad oblast, Russian Federation.

3) Department of Physics, St. Petersburg State University, 198504, St. Petersburg, Russian Federation.

**Abstract**. Carbon dioxide fixation and storage constitute a drastically important problem for the humanity nowadays. We hereby publish a new solution based on the alkaline earth salts with a weakly coordinating anion, tetrakis(pentafluorophenyl)borate. The proposed solution was validated using a robust combination of global minimum search and molecular dynamics simulations utilizing a well-tested, reliable semiempirical Hamiltonian to monitor chemical reactions. Calcium tetrakis(pentafluorophenyl)borate captures 5.5 $CO_2$ molecules per mole, whereas barium tetrakis(pentafluorophenyl)borate captures 3.6 $CO_2$ molecules per mole. These capacities are much higher, as compared to the established carbonate technology, which fixes only one $CO_2$ molecule per one metal atom. The conducted simulations reveal that electrostatic binding of $CO_2$ to alkaline earth cations is more technologically interesting than formation of carbonate salts. Our simulation results can be directly validated by sorption measurements.

**Key words:** carbon dioxide capture; carbonate looping; alkaline earth; tetrakis(pentafluorophenyl)borate; PM7-MD.

---

[A] Corresponding author: vvchaban@gmail.com (Prof. Vitaly V. Chaban).

**TOC Image**

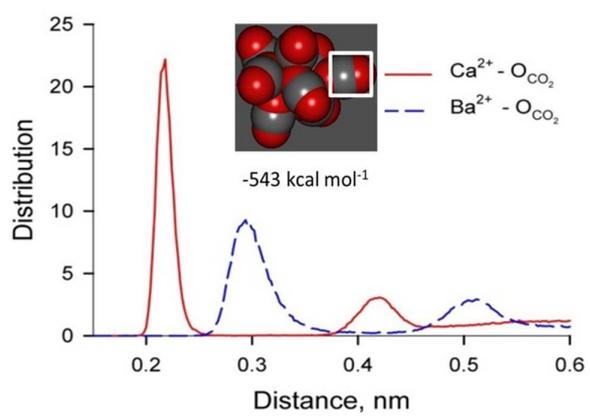

**Introduction**

Long-term climate changes and continuous global warming constitute a serious ecological challenge for the humanity nowadays.[1] A gradual elevation of the greenhouse gas concentrations, the major one of which is carbon dioxide ($CO_2$), is linked to the higher average temperatures throughout the globe. Industrial activities of humanity – such as combustion of coal, petroleum, and natural gas – inevitably foster higher emission levels of $CO_2$ and other greenhouse gases. Efficient removal of $CO_2$ upon combustion of the fossil fuels and its further storage remain to be an urgent problem.[2,3]

Amine scrubbing is presently the most wide-spread method to capture $CO_2$ in the industrial conditions, established over the past 60 years.[4-6] It performs generally better, as compared to ionic liquids,[7-12] polymeric membranes,[13-15] cryogenic distillation, gas hydrates, and less common technologies.[16-19] In the meantime, the amine scrubbing technology suffers from a few principal drawbacks: (1) a relatively small number of absorption cycles due to solvent losses, (2) high regeneration costs, and (3) ecological consequences. Development of new $CO_2$ scavengers with the primary purpose to simultaneously decrease volatility of the sorbent and regeneration energy is underway.

The carbonate (solid salts at room conditions) formation upon reaction of basic and acidic oxides is well-known. $CO_2$ belongs to acidic oxides. Thus, it is possible to fix $CO_2$ by the basic oxides, most of which are readily decomposable upon some heating. This technology has reached a rather established state nowadays, being known as carbonate looping or, in the case of calcium oxide, calcium looping.[20-26] The carbonation reaction occurs both at the solid-gas interface and in the aqueous solutions offering possibilities for tunable technological setups. In water, the carbonates are transformed into bicarbonates assuming an excess of supplied $CO_2$. Therefore, one alkali metal ion keeps one $CO_2$ molecule, whilst one alkaline earth metal ion keeps two $CO_2$ molecules in the form of bicarbonate. The above numbers constitute theoretical maxima. Unfortunately, alkali metal bicarbonates are soluble in water, liberating unstable $H_2CO_3$. Thus, $CO_2$ is essentially returned back to the atmosphere. Dry alkali metal carbonates thermally decompose between 100-200 ºC releasing $CO_2$ and water vapors, which must be subsequently condensed.[27]

In addition to the main-group metals, most transition metal oxides, e.g. manganese, iron, cobalt, nickel, copper, zinc etc, can be easily carbonated from a thermodynamic point of view. However, using the above oxides for the $CO_2$ capture applications is impractical due to the importance of the corresponding metals for other modern technologies and their precious features. Significant amounts of $CO_2$ to be stored require large amounts of raw materials as a feedstock for carbonation. An ideal metal must be abundant, cheap and give rise to carbonates with a limited aqueous solubility.

Development of new metal containing compounds with specifically elaborated structures for $CO_2$ capture is actively pursued. Duan and co-authors combined density functional theory (DFT) and lattice phonon dynamics to provide a quantitative comparison of the different alkaline earth metal oxides in the context of their carbonation. Interestingly, the MgO and $Mg(OH)_2$ samples were found to perform best for $CO_2$ fixation, in particular, due to their lower operating temperatures (600-700 K).[28] Kim and co-authors conducted first-principles calculations to rate metal promoters of MgO sorbents. Five alkali (lithium to cesium) and four alkaline earth metals (beryllium to barium) were screened. In comparison with the $CO_2$ absorption energy by pure MgO, the absorption energy on the metal-promoted MgO material appears higher. These results encourage to apply metal promotion to enhance sorption capacities of magnesium oxide.[29]

Naturally occurring Ca-based materials are inexpensive, but exhibit a very rapid decay of the $CO_2$ capacity after a few cycles, since the involved chemical reactions are not ideally reversible. Recall that a limited number of working cycles is also an essential problem in the context of amine scrubbing. Recently, Broda and co-authors developed synthetic Ca-based $CO_2$ sorbents using a sol-gel technique.[30] The nanostructured material possesses a high surface area and pore volume and exhibits an acceptable $CO_2$ capacity over numerous cycles. Ca-decorated fullerene ($Ca@C_{60}$) for $CO_2$ and $N_2O$ adsorption was theoretically characterized by Gao and co-authors using DFT.[18] $CO_2$ and $N_2O$ adsorptions at $Ca@C_{60}$ are much enhanced, as compared to pristine $C_{60}$. Up to 5 $CO_2$ molecules are fixed by one $Ca@C_{60}$, thanks to the electrostatic interaction, which can also be referred to as ionic bonding. Since an ionic bond is non-directional and unsaturated, the maximum number of the $CO_2$ molecules attracted depends only on the space available around the cation. Note that the exemplified process does not belong to carbonate looping, since no formal chemical reaction takes place. The theoretically proposed hybrid material remains to be constructed in practice and the problem of its dispersion over some $CO_2$-philic media remains to be solved.

To recapitulate, the carbonate looping technology allows to store 0.5 to 1 $CO_2$ molecules per one metal atom in the form of carbonate, depending on the valence of the latter. After being converted into carbonate, the $CO_2$ capturing ability of the metal decreases, since the electrostatic attraction of oxygen ($CO_2$) to the cation is strongly screened by the surrounding carbonate anions. Furthermore, inorganic carbonates are solid substances. Thus, crystallization occurs quickly, prohibiting penetration of $CO_2$ inside the sorbent. In the absence of moisture, however, a possibility exists to avoid carbonate formation and keep $CO_2$ capture dependent only on the ion-molecular attraction. The maximum number of captured $CO_2$ in the latter case depends on the volume available around the cation and the cation-$CO_2$ binding energy. Since the cation–$CO_2$ binding is expectedly weaker than the cation–$CO_3^{2-}$ binding, a regeneration energy cost should become smaller. To corroborate the above idea, we hereby systematically investigated $CO_2$ capture by the calcium and barium cations, supplied in the form of salts employing one of the most weakly coordinating anion, tetrakis(pentafluorophenyl)borate ($TFPB^-$). $TFPB^-$ is a bulky anion exhibiting significant flexibility and highly delocalized excess electronic charge. Free alkaline earth cations were also simulated for reference.

**Methodology**

Initial geometries of the simulated systems containing single cations ($Ca^{2+}$ and $Ba^{2+}$), the counter-ions ($Cl^-$, $TFPB^-$), and variable number of $CO_2$ and $H_2O$ molecules (4 to 25) were constructed in PACKMOL[31] using multiple arbitrary rotations of the ions and molecules to minimize starting potential energy. Subsequently, the geometries of the systems were optimized by the eigenfollowing algorithm and the PM7 semiempirical Hamiltonian.[32]

PM7 is based on the Neglect of Diatomic Differential Overlap (NDDO) approximation. Introduction of NDDO has two purposes. First, the wave function convergence is made faster, as compared to Hartree-Fock calculations. Second, the parameters, which substitute certain integrals, increase an accuracy of the method, since they are derived from experimental and high-level ab initio reference data. Therefore, many quantum effects, unavailable to the conventional Hartree-Fock calculations, are effectively present in the PM7 calculations. The PM7 Hamiltonian contains specific modifications to improve representation of the non-bonded interactions. In particular, in-built corrections for hydrogen bonding, peptide bonding, halogen bonding, and dispersion attraction are routinely applied in all simulated systems.[33-38] An accuracy of PM7 for non-bonded interactions was recently investigated, having provided encouraging results.[39] The accuracies of PM7, as applied to hundreds of very different systems, were reported elsewhere.[35,36,40]

The global minimum configurations of the ion-molecular systems were obtained by the annealing procedure, as previously exemplified by us in Ref.[41] The ion-molecular motion was simulated at 2000 K during 5.0 ps with a time-step of 0.1 fs. The temperature of the system was coupled to the Andersen thermostat.[42] Every 0.1 ps (1000 time-steps), immediate coordinates of all atoms were saved. Thus, 50 supposedly very different starting ion-molecular configurations were obtained for every system. The forces acting on every atom of the system were decreased down to 4.2 kJ mol$^{-1}$ Å$^{-1}$ (gradient norm convergence threshold) by the eigenfollowing algorithm. The configuration exhibiting the smallest standard heat of formation was assumed to be the global-minimum one. The frequency analysis was performed to ensure that no negative vibrational frequencies are present.

Coulson charge fluctuations and radial distribution functions were obtained from PM7-MD simulations at 300 K. The trajectory length was 200 ps with a time-step of 0.5 fs. PM7-MD follows Born-Oppenheimer approximation. The nuclei moved in time based on the immediate forces according to the PM7 Hamiltonian. Valence electrons were treated explicitly, whereas an effect of core electrons was provided by pseudopotentials to enhance convergence. The wave function convergence threshold was set to 10$^{-5}$ Hartree. PM7-MD was previously applied with success to versatile problems in chemistry having provided satisfactory to excellent results.[41,43-46] based on our previous studies and benchmarks.

MOPAC ver. 2012 (openmopac.net) was used to conduct electronic-structure calculations and apply the eigenfollowing algorithm. VMD (ver. 1.9.1)[47] and Gabedit (ver. 2.8)[48] were used to visualize systems and prepare molecular artwork.

**Results and Discussion**

In the sake of Hamiltonian and method validation, we started with the simulation of the classic CaO + CO$_2$ = CaCO$_3$ reaction in the gas phase using 4 CO$_2$ molecules and 4 CaO particles. Whereas calcium oxide exists as a crystalline solid at standard conditions, representation of it in the form of CaO molecules is acceptable as long as all reactive sites are available to CO$_2$. It was found after moderate sampling that 3 CO$_2$ molecules readily produced 3 carbonate anions, which formed a tiny subnanoparticle of CaCO$_3$ (Figure 3). Therefore, PM7 correctly represents immediate forces in the system and correctly predicts the experimentally known product. Annealing at 2000 K is essential to overcome potential energy barriers. All $Ca^{2+}$ ions were incorporated into the nascent structure of $CaCO_3$. One $CO_2$ molecule in the simulated system, however, remained intact. More extensive sampling is required to observe this last transformation of CaO into $CaCO_3$. Note that reaction pace decreases substantially with the decrease of reactant concentrations, therefore, the real-time simulation up to the very final state may be quite time consuming.

Capturing $CO_2$ by $Ca(CO_2)^{2+}$ is 1.6 times more thermodynamically favorable than that by $Ca(CO_3)^0$. In terms of binding energy, the attachment of the second $CO_2$ molecule to $Ca^{2+}$ brings -201 kJ mol$^{-1}$, whilst its attachment to $CaCO_3$ brings only -126 kJ mol$^{-1}$. Nonetheless, the major reason of inability of $CaCO_3$ to fix additional gas molecules is its crystallization, which makes most newly formed $CaCO_3$ entities unavailable to external $CO_2$.

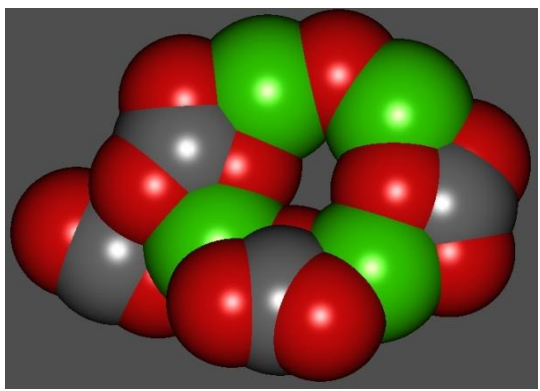

Figure 1. Atomistic configuration obtained by the global energy minimum search in the system "4 CaO + 4 $CO_2$". Color codes: calcium is green, oxygen is red, carbon is grey.

$Ca^{2+}$ coordinates up to 8 $CO_2$ molecules in the first sphere (Figure 2) forming multiple well-defined calcium-oxygen contacts. Each additional $CO_2$ molecule goes to the second coordination sphere. Small size of $CO_2$ and its linear structure favor the compact packing of molecules around the cation and the large coordination number. In the absence of water or other oxygen source, $CO_2$ does not form the carbonate anion. Therefore, gaseous $CO_2$ alone is principally unable to neutralize $Ca^{2+}$ by giving rise to $CaCO_3$. A purely ionic bonding between $Ca^{2+}$ and $CO_2$ in the absence of water is highly beneficial in the context of $CO_2$ capture, since this process increases the number of fixed $CO_2$ by 8 times, as compared to conventional carbonate looping. The energetic gain, in terms of standard heat of formation (Figure 3), of $CO_2$ coordination decreases slightly as new molecules join the first coordination sphere. For instance, the capture of the fifth $CO_2$ molecule by the $Ca(CO_2)_4^{2+}$ cluster brings -110 kJ mol$^{-1}$, whereas the capture of the eighth $CO_2$ molecule by the $Ca(CO_2)_7^{2+}$ cluster brings -103 kJ mol$^{-1}$. In turn, the ninth $CO_2$ molecule, which joins the second coordination sphere, brings only -93 kJ mol$^{-1}$, since its attraction to $Ca^{2+}$ decreases due to the larger interatomic distance. Noteworthy, all $CO_2$ molecules in the first coordination sphere exhibit the O-C-O valence angle of 177°. In turn, the same angle in the ninth molecules equals to 180°. The three-degree alteration of the valence angle suggests a strong non-covalent interaction between the doubly charged cation and the neighboring gas molecules.

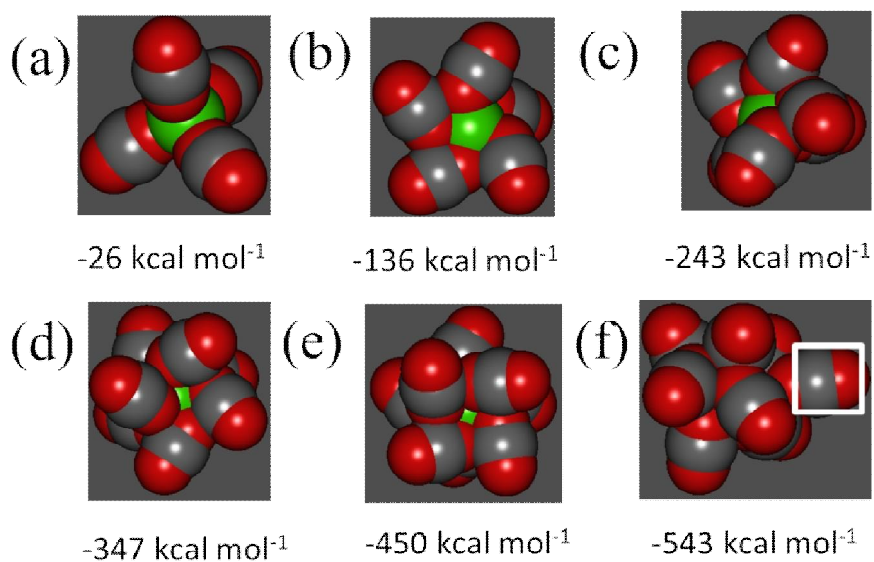

Figure 2. Global-minimum ion-molecular configurations: (a) $Ca(CO_2)_4^{2+}$, (b) $Ca(CO_2)_5^{2+}$, (c) $Ca(CO_2)_6^{2+}$, (d) $Ca(CO_2)_7^{2+}$, (e) $Ca(CO_2)_8^{2+}$, (f) $Ca(CO_2)_9^{2+}$. The white frame designates $CO_2$ molecule belonging to the second coordination sphere of $Ca^{2+}$. The depicted numbers correspond to the standard heats of formation of the respective structures. Color codes: oxygen is red, carbon is grey, calcium is green.

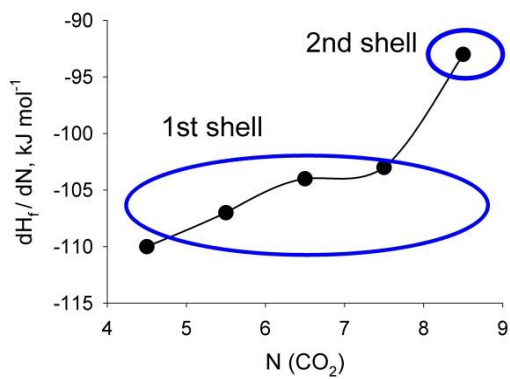

Figure 3. Derivatives, $dH_f/dN$, of the standard heat of formation with respect to the number of the $CO_2$ molecules in $Ca(CO_2)_n$. The number of provided $CO_2$ molecules corresponds to the first and the second solvation shells of the calcium cation.

The average calcium-oxygen distance in the simulated $Ca(CO_2)_N^{2+}$ clusters depends somewhat on the number of captured $CO_2$ molecules (Figure 4). In $Ca(CO_2)_4^{2+}$, the $CO_2$ molecules are closer to $Ca^{2+}$, the average Ca-O distance is 2.07 Å. In $Ca(CO_2)_8^{2+}$, the distance is by 0.11 Å larger. Due to competition of the $CO_2$ molecules to coordinate $Ca^{2+}$ in the crowded first coordination sphere and the respective steric hindrance, an average distance and, therefore, electrostatic attraction decays insignificantly, but systematically. The same effect was recorded by us upon simulation of other polar molecules, such as water, hydrogen fluoride, acetonitrile, and alcohols. The binding energy in $Ca(CO_2)^{2+}$ amounts to ~170 kJ mol$^{-1}$, which much exceeds the thermal energy per degree of freedom and makes $CO_2$ capture a highly thermodynamically favorable physical process. Some decrease of the energetics of this process is even beneficial to decrease regeneration costs of the proposed $CO_2$ scavenger.

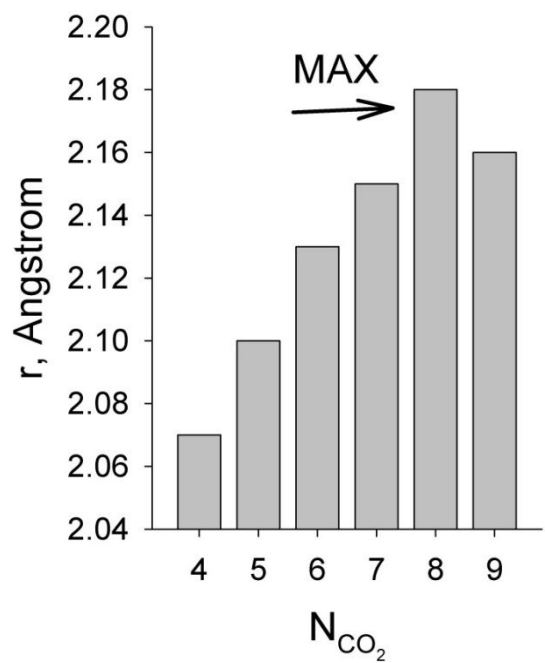

Figure 4. Average Ca-O distance in the first coordination sphere of $Ca^{2+}$ in the $Ca(CO_2)_N^{2+}$ clusters.

It was shown above that $Ca^{2+}$ captures a number of $CO_2$ molecules in vacuum and the underlying reactions are thermodynamically favorable. The recorded thermodynamic gains are significant, exceeding the $k_B \times T$ product per degree of freedom at room conditions. Thanks to the ability of $Ca^{2+}$ to keep 8 $CO_2$ molecules in the first coordination sphere, this scavenger appears to be a few times more efficient than CaO in the carbonate looping technology. In real salts, $Ca^{2+}$ always coexists with anions. The anions differ very substantially based on their coordination ability. The goal of molecular design is to find such anions, which do not block $CO_2$ approaching to the cation and its subsequent sorption. In the following, we consider an effect of the strongly non-coordinating anion (TFPB$^-$). TFPB$^-$ is a bulky, massive, and chemically inert particle, whose excess electronic charge is delocalized over the entire volume of the anion. The cation-anion interactions in such salts are relatively weak, since the anion does not possess a single strongly charged site. Many sites of the anion compete simultaneously for the most energetically favorable position near $Ca^{2+}$. The intermolecular structure of the TFPB$^-$–based salts is relatively weakly defined (Figure 5), that may result in a low melting temperature. There is no experimental information on [Ca][TFPB] thus far, but it can be supposedly synthesized in analogy with [Li][TFPB].

Lone cations – $Ca^{2+}$ and $Ba^{2+}$ – and optimized geometries of their salts – [Ca][TFPB] and [Ba][TFPB] – were used to generate four MD systems, containing 25 $CO_2$ molecules each. These MD systems are not periodic (clusters-in-vacuum) and the genuine dynamics of atoms in these systems is very quick at 300 K. Due to the absence of the closest images, the long-range structure of the systems is essentially neglected in the reported PM7-MD simulations. This simplification should have a negligible impact on the sorption processes, since $CO_2$ is a gas at room conditions.

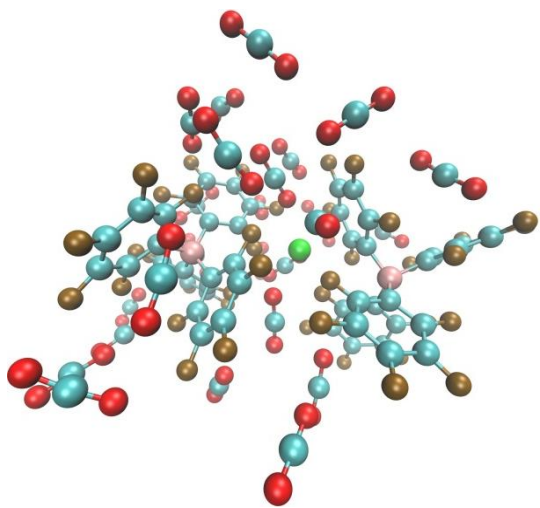

Figure 5. An immediate ionic configuration of the "$Ca^{2+}$ + 2 TFPB$^-$ + 25 $CO_2$" system, which was recorded after a thermodynamic equilibrium was reached. Calcium atom is green, boron atoms are pink, carbon atoms are cyan, fluorine atoms are brown, and oxygen atoms are red. PM7-MD simulation was conducted at 300 K.

Heat of formation of the solvated $Ca^{2+}$ and $Ba^{2+}$ and their solvated salts fluctuates insignificantly and with high frequency. The equilibrium states are reached very quickly, within ~10 ps (lone cations) and ~20 ps (salts) due to the absence of the long-range order in these systems. The systems containing calcium exhibit somewhat lower heats of formation (-8.0 vs. -7.8 MJ mol$^{-1}$), indicating their higher thermodynamic stabilities. The introduced thermal energy (300 K) is used largely by the $CO_2$ molecules to overcome potential energy barriers and find their most energetically favorable positions around the cations. The employed alkaline earth cations offer a high density of electronic charge on their surface. $CO_2$ does not possess a dipole moment, but instead possesses two polar covalent bonds, resulting in certain partial charges on every atom. Electrostatic attraction of the oxygen atoms to the cation is stronger than the repulsion of somewhat electron deficient carbon atom. The number of $CO_2$ molecules, which $Ca^{2+}$ and $Ba^{2+}$ accommodate, is determined by the space available around their surfaces. One would assume that the roles of the small and large anions, as well as the role of the cation-anion binding, are different in this case.

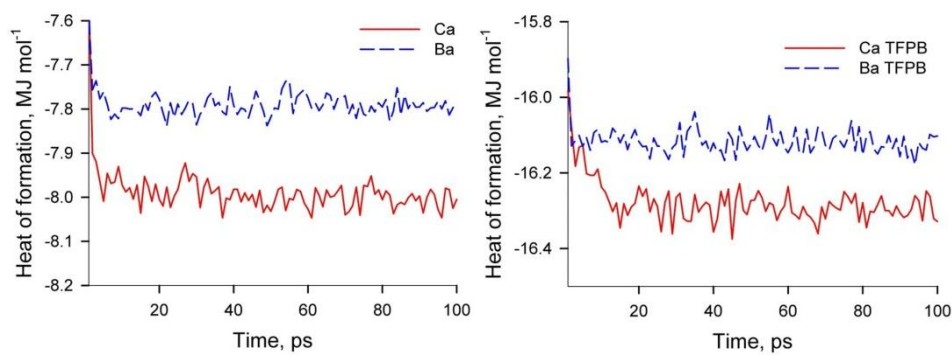

Figure 6. Fluctuations of the standard heat of formation during the first 200,000 time-steps at finite temperature, 300 K, for lone ion (left) and ion pair (right) in the course of PM7-MD simulations (see legend on figure). Hereby, the first 100 ps are shown to display equilibration parts of the trajectories, whereas the total length of each trajectory is 200 ps.

Radial distribution function (RDF) is a straightforward instrument to systematically characterize all possible distances in the atomistic system. RDFs for the cation-oxygen ($CO_2$) distance distributions for the lone cations – $Ca^{2+}$, $Ba^{2+}$ – and their salts – [Ca][TFPB] and [Ba][TFPB] – are given in Figure 7. Since the calcium atom is smaller than the barium atom, its minimum distance to $CO_2$ is also somewhat smaller, compare 0.22 nm to 0.30 nm. Both distances indicate strong electrostatic coordination, but exclude covalent bonding, as it would occur in the case of the carbonate and hydrocarbonate anions. The second peaks (Figure 7) are observed at 0.41 and 0.51 nm, corresponding to the second (more distant) oxygen atom of the same coordinated $CO_2$ molecule. No second coordination spheres are present in these systems at 300 K. Importantly, the positions of both peaks do not depend on the presence of the anion. Their heights only moderately decrease when the anion is introduced. The peak for barium-oxygen is 2.5 times smaller than that for calcium-oxygen. This is explained by a lower charge density on $Ba^{2+}$, due its larger volume, whereas the formal electrostatic charge is the same for all alkaline earth cations.

Integration of the RDFs up to the first minimum provides coordination numbers, characterizing the first coordination spheres, formed around the alkaline earth cations. In particular, the coordination number at 300 K is 7.7 for lone $Ca^{2+}$ and 3.9 for lone $Ba^{2+}$. Recall, the coordination number of $Ca^{2+}$ at zero temperature is slightly larger, amounting to 8. The anions (2×TFPB$^-$) push some $CO_2$ molecules from the first coordination sphere of the cations, providing thereby 5.5 for $Ca^{2+}$ and 3.6 for $Ba^{2+}$. Thus, the role of the weakly coordinating anion is much larger for the calcium salt.

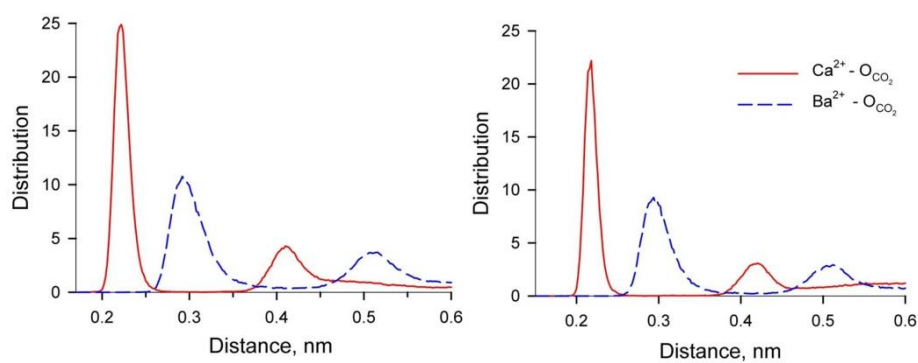

Figure 7. Distance distribution functions computed for the alkaline earth cation and the oxygen atom of $CO_2$: $Ca^{2+}$ (red solid line) and $Ba^{2+}$ (blue dashed line) in the case of the lone cation (left) and the cation-anion pair (right). The anion chosen is TFPB$^-$.

Partial charges, evolution of which during the first 100 ps of the MD simulations is provided in Figure 8, quantify a stronger coupling between $Ca^{2+}$ and $CO_2$. Whereas no covalent bond is formed in the complex, $Ca^{2+}$ gets 0.6e from several neighboring $CO_2$ molecules. In turn,

Ba$^{2+}$ gets only 0.2e. In other words, the CO$_2$ molecules in the first coordination sphere are slightly electron deficient. The fractions of the shared electron density can be understood as a qualitative measure of the degree of covalence in the ionic bond, and this fraction is larger in the case of Ca$^{2+}$. Smaller partial charges were also previously reported for lighter and, therefore, smaller alkali cations in their complexes with counterions and neutral particles.[49]

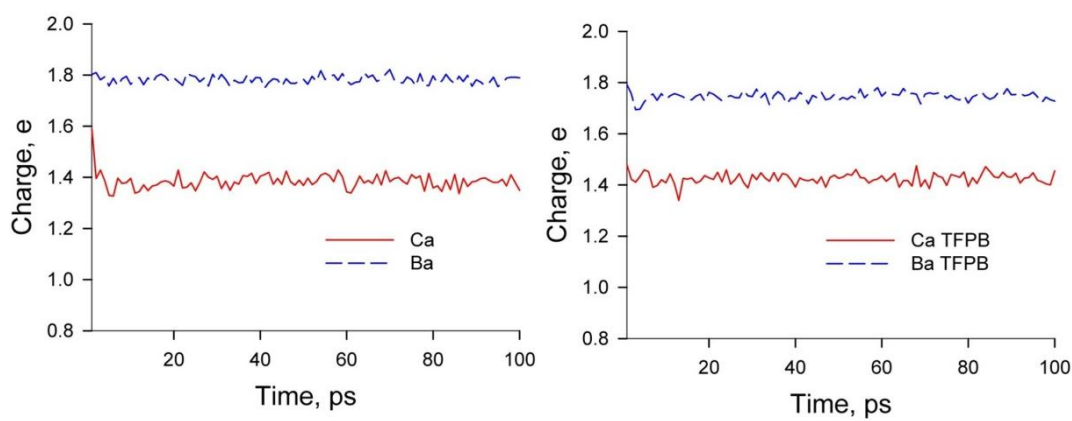

Figure 8. Coulson partial charge distributions for lone cations Ca$^{2+}$/Ba$^{2+}$ (left) and ion pairs [Ca][TFPB]/[Ba][TFPB] (right) during PM7-MD simulations.

**Conclusions**

PM7-powered global minimum search and PM7-MD simulations were hereby conducted to validate the concept of CO$_2$ capture by the alkaline earth metal salts. This new solution for the CO$_2$ capture falls into contrast with the well-known carbonate looping, in which the carbonate salt is formed from the metal oxide and CO$_2$. In the carbonate looping, the cation is neutralized by the carbonate anion making further capture of the CO$_2$ molecules less efficient. We found that a lone calcium cation keeps up to 8 CO$_2$ molecules in its first coordination sphere assuming no competing coordination. Therefore, the CO$_2$ capacity can be drastically increased by blocking or hindering the cation-anion coordination.

In the proposed solution, $Ca^{2+}$ was paired with a very weakly coordinating singly charged tetrakis(pentafluorophenyl)borate anion. Two such anions are needed to neutralize $Ca^{2+}$. Despite a huge size, these anions only insignificantly modify the first coordination sphere of lone $Ca^{2+}$, decreasing the total number of the coordinated $CO_2$ molecules down to 6. Thus, the $CO_2$ capacity of $Ca^{2+}$ can be increased by 6 times if a suitable counter-ion is chosen. It is a sound expectation that weakly coordinating anions undermine structure of the salt and introduce a number of 'pockets', which can be impregnated by small molecule, such as $CO_2$. Furthermore, electrostatic attraction between $CO_2$ and the cations is more preferable in the context of regeneration, since less energy is required to break such bonding. Experiments are urged to characterize a regeneration performance of this new $CO_2$ scavenger.

$Ba^{2+}$ is larger than $Ca^{2+}$ and, therefore, $CO_2$ molecules are generally located farther from it and the corresponding ion-dipole attraction is weaker. The computed partial Coulson charge of $Ca^{2+}$ amounts to +1.4e, whereas the charge of $Ba^{2+}$ is +1.8e. However, considering the van der Waals volume of these alkaline earth ions, one concludes that the deficient electron charge on $Ba^{2+}$ is more delocalized. This observation is in concordance with the poorer $CO_2$ capturing ability of $Ba^{2+}$: 3.9 by the lone cation and 3.6 by [Ba][TFPB].

To recapitulate, a new $CO_2$ capturing method has been proposed and validated by quantum chemical simulations. Experimental efforts are now urged to synthesize [Ca][TFPB] and [Ba][TFPB] to assess interaction of their bulk phases with $CO_2$. Future research efforts must be directed towards investigation of these salts structures, their phase transition points, and an effect of the absorbed $CO_2$ on their physicochemical properties.

**Author Information**

E-mail for correspondence: vvchaban@gmail.com (V.V.C.)